\documentclass[runningheads,a4paper]{llncs}

\makeatletter
\newcommand{\printfnsymbol}[1]{%
  \textsuperscript{\@fnsymbol{#1}}%
}
\makeatother

\usepackage{amssymb}
\usepackage{amsmath}
\usepackage{graphicx}

\usepackage{url}
\usepackage[breaklinks=true,letterpaper=true,colorlinks,bookmarks=false]{hyperref}
\usepackage{xcolor}
\urldef{\mailsa}\path|yuxing.tang@nih.gov, rms@nih.gov|
\usepackage{diagbox}
\usepackage{multirow}
\usepackage{stmaryrd}
\usepackage{bbm}
\newcommand{\etal}{\textit{et al}.}
\newcommand{\eg}{\textit{e}.\textit{g}.}
\newcommand{\norm}[1]{\lVert#1\rVert}

\begin{document}

\mainmatter  % start of an individual contribution

\title{TUNA-Net: Task-oriented UNsupervised Adversarial Network for Disease Recognition in Cross-Domain Chest X-rays}
\titlerunning{TUNA-Net for Cross-Domain Disease Recognition}

\author{Yuxing~Tang$^1$,
Youbao~Tang$^1$,
Veit~Sandfort$^1$,
Jing~Xiao$^2$,
and Ronald~M.~Summers$^1$}
% index{Tang, Yuxing}
% index{Tang, Youbao}
% index{Sandfort, Veit}
% index{Xiao, Jing}
% index{Summers, Ronald}

\authorrunning{Y.X.~Tang, Y.B.~Tang, V.~Sandfort and R.M.~Summers}
\institute{% Department of Radiology and Imaging Science, Clinical Center,\\
 $^1$National Institutes of Health, Clinical Center, Bethesda, MD, USA\\
 \mailsa \\
 $^2$Ping An Technology Co., Ltd., Shenzhen, China}

\maketitle

\begin{abstract}
In this work, we exploit the unsupervised domain adaptation problem for radiology image interpretation across domains. Specifically, we study how to adapt the disease recognition model from a labeled source domain to an unlabeled target domain, so as to reduce the effort of labeling each new dataset. To address the shortcoming of cross-domain, unpaired image-to-image translation methods which typically ignore class-specific semantics, we propose a task-driven, discriminatively trained, cycle-consistent generative adversarial network, termed TUNA-Net. It is able to preserve 1) low-level details,  2) high-level semantic information and 3) mid-level feature representation during the image-to-image translation process, to favor the target disease recognition task. The TUNA-Net framework is general and can be readily adapted to other learning tasks. We evaluate the proposed framework on two public chest X-ray datasets for pneumonia recognition. The TUNA-Net model can adapt labeled adult chest X-rays in the source domain such that they appear as if they were drawn from pediatric X-rays in the unlabeled target domain, while preserving the disease semantics. Extensive experiments show the superiority of the proposed method as compared to state-of-the-art unsupervised domain adaptation approaches. Notably, TUNA-Net achieves an AUC of 96.3\% for pediatric pneumonia classification, which is very close to that of the supervised approach (98.1\%), but without the need for labels on the target domain.
\end{abstract}

\section{Introduction}
While deep convolutional neural networks (CNNs) have achieved encouraging results across a number of tasks in the medical imaging domain, they frequently suffer from generalization issues due to source and target domain divergence. Examples of such divergence include distribution shift caused by images collected with distinct protocols, from different institutions or patient groups. This can be alleviated by \textit{supervised domain adaptation} (SDA)~\cite{DA_MRI_MICCAI17,Zhang_2018_CVPR}, which adapts certain layers of the model that trained with large amounts of well-labeled source data, with additional moderate amounts of labeled target data. However, obtaining abundant labels in each new, unseen domain is a non-trivial and laborious process that relies heavily on skilled clinicians in the majority of clinical applications. Alternatively, \textit{unsupervised domain adaptation} (UDA)~\cite{ADDA_CVPR17} aims to mitigate the harmful effects of domain divergence when transferring knowledge~\cite{Tang_2016_CVPR,Tang_2017_TPAMI} from a supervised (labeled) source domain to an unsupervised (unlabeled) target domain. Because of its potential benefits for medical image processing, UDA of deep learning models has attracted many researchers' attention~\cite{UDA_IPMI17,TDGAN_MICCAI18,SeUDA_MLMI18}.

Adversarial adaptation methods~\cite{ADDA_CVPR17,Cycada2017} have become increasingly popular with the recent success of generative adversarial networks (GANs)~\cite{GAN} and their variants~\cite{CycleGAN2017}. In medical imaging, most of the previous work for adversarial adaptation focuses on lesion or organ segmentation~\cite{UDA_IPMI17,Zhang_2018_CVPR,TDGAN_MICCAI18,SeUDA_MLMI18}. For instance, Kamnitsas \etal~\cite{UDA_IPMI17} derive domain-invariant features by an adversarial network for brain lesion segmentation of MR images from two different datasets. GAN-based image-to-image (I2I) translation methods~\cite{CycleGAN2017} are also widely used to generate medical images~\cite{tang2019xlsor,Tang_2019_ISBI} cross modalities to help adaptation. For example, Zhang \etal~\cite{TDGAN_MICCAI18} segment multiple organs in unlabeled X-ray images with labeled digitally reconstructed radiographs rendered from 3D CT volumes, using I2I translation. Zhang \etal~\cite{Zhang_2018_CVPR} improve Cycle-GAN~\cite{CycleGAN2017} by introducing shape-consistency for CT and MRI cardiovascular 3D image translation to help organ segmentation. Though CT and MR images are not necessarily paired, the shape-consistency loss requires supervision of pixel-wise annotations from both domains. Chen \etal~\cite{SeUDA_MLMI18} preserve semantic structural information of the lungs in chest radiographs (X-rays) for cross-dataset lung segmentation.

All the previous methods deal with limited domain shift or large organs appearing at approximately fixed positions with clear boundaries, or both. Moreover, they do not necessarily preserve \textbf{class-specific} semantic information of lesions or abnormalities in the process of distribution alignment. An illustrative example is, when translating an adult X-ray into a pediatric X-ray, there is no guarantee that fine-grained disease content on the original image will be explicitly transferred. The capability of preserving class-specific semantic context across domains is crucial for medical imaging analysis for certain clinically relevant tasks, such as disease or lesion classification, detection and segmentation~\cite{Tang_MLMI2018,Tang_SPIE2019,TangYB_2019_ISBI,yan2019mulan}.  However, to our best knowledge, solutions to this problem of adversarial adaptation for medical imaging are limited.

In this paper, we present a novel framework to tackle the target task of disease recognition in cross-domain chest X-rays. Specifically, we proposed a task-oriented unsupervised adversarial network (TUNA-Net) for pneumonia (findings on X-rays are airspace opacity, lobar consolidation, or interstitial opacity) recognition in cross-domain X-rays. Two visually discrepant but intrinsically related domains are involved: adult and pediatric chest X-rays. The TUNA-Net consists of a cyclic I2I translation framework with class-aware semantic constraint modules. In the absence of labels from one domain, the proposed model is able to 1) synthesize ``radio-realistic'' (i.e., a synthesized radiograph that appears anatomically realistic) images with sufficient low-level details across two different domains, 2) preserve high-level class-specific semantic contextual information during translation, 3) regularize learned mid-level features of real and synthetic target domains to be similar, 4) optimize the objective functions simultaneously to generalize to the unlabeled domain. We demonstrate the effectiveness of our approach on two public chest X-ray datasets of sufficient domain shift for pneumonia recognition. 

\section{Method}

\subsection{Problem Formulation}
In this work, we focus on the problem of unsupervised domain adaptation, where we are given a source domain $A$ with both images $X_A$ (\eg, adult X-rays) and labels $Y_A$ (\eg, normal or pneumonia), and a target domain $P$ with only images $X_P$ (\eg, pediatric X-rays), but no labels. The goal is to learn a classification model $\mathcal{F}$ from images of both domains but with only source labels and predict the labels in the target domain. Note that $X_A$ are naturally unpaired with $Y_P$ as these images are from two different patient populations (adults and children).

\begin{figure*}[ht]
	\centering
	\includegraphics[width=\linewidth]{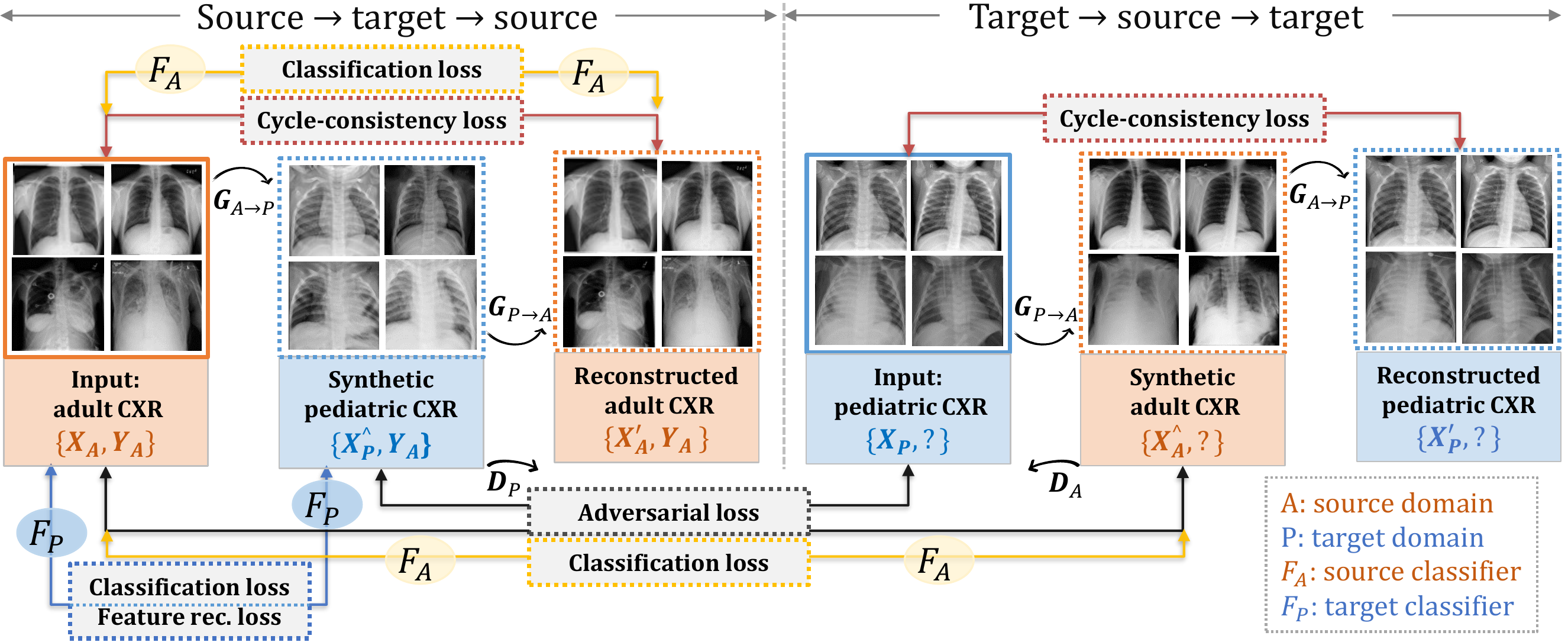}
	\vspace{-6mm}
	\caption{The framework of TUNA-Net. The question we investigate is whether class-specific semantics can be preserved in an I2I translation framework (\eg, Cycle-GAN~\cite{CycleGAN2017}) to help domain adaptation, providing disease labels only in source domain (\eg, translate an adult chest X-ray into a pediatric chest X-ray while preserving disease semantics, i.e., normal or pneumonia). In test phase, model $\mathcal{F}_P$ is applied on target pediatric images to make predictions. In this figure, for inputs from both domains, top two examples are normal, bottom two examples are with pneumonia.}
	\label{fig:overview}
\end{figure*}

A naive baseline method is to learn $\mathcal{F}$ solely from source images and labels, then apply it directly on target domain. While $\mathcal{F}$ performs well on data with similar distribution as the source data, it typically leads to degraded performance on the target data because of domain divergence.
To alleviate this effect, we follow previous methods~\cite{CycleGAN2017,Zhang_2018_CVPR,TDGAN_MICCAI18} to map images from two domains ($X_A\leftrightarrows{X_P}$) using multi-domain I2I translation with unpaired training data. During translation, we add constraints at different levels to preserve both holistic and fine-grained class-specific image content. Consequently, the model $\mathcal{F}$ learned on the source domain can be well generalized to the target domain. The flowchart of the proposed framework for UDA is shown in Figure~\ref{fig:overview}.

%%%%%%%%%%%%%%%%%%%%%%%%%%%%%%%Fig%%%%%%%%%%%%%%%%%%%%%%%%%%%%%%%%%%%%%%%%%%%%%%%%%%%
%%%%%%%%%%%%%%%%%%%%%%%%%%%%%%%%%%%%%%%%%%%%%%%%%%%%%%%%%%%%%%%%%%%%%%%%%%%%%%%%%%%%%%

\subsection{Pixel-level image-to-image translation with unpaired images}

GANs~\cite{GAN} have been widely used for image-to-image translation. Given unpaired images from two domains, we adopt Cycle-GAN~\cite{CycleGAN2017} %(or similar method such as ~\cite{MUNIT}) 
to first learn two mappings: $A\shortrightarrow P$ and $P\shortrightarrow A$, with two generators $G_{A\shortrightarrow P}(X_A)$ and $G_{P\shortrightarrow A}(X_P)$, so that discriminators $D_P$ and $D_A$ can not distinguish between real and synthetic images generated by $G$. For $G_{A\shortrightarrow P}$ and its discriminator $D_P$, the objective is expressed as the \textit{adversarial learning loss}:

\begin{equation}
    \mathcal{L}_{\text{adv}}(G_{A\shortrightarrow{P}},D_P) = \mathbb{E}_{x_a \sim X_A}[\log (1-D_P(G_{A\shortrightarrow P}(x_a))] + \mathbb{E}_{x_p \sim X_P}[\log D_P(x_p)].
\end{equation}

A similar adversarial loss can be designed for mapping $G_{P\shortrightarrow A}$ and its discriminator $D_A$ as well: i.e., $\min_{G_{P\shortrightarrow{A}}}\max_{D_A}\mathcal{L}_{\text{adv}}(G_{P\shortrightarrow{A}},D_A)$.

To preserve sufficient low-level content information for domain adaptation, we then use the \textit{cycle consistency loss}~\cite{CycleGAN2017} to force the reconstructed synthetic images $x_a'$ and $x_p'$ to resemble their inputs $x_a$ and $x_p$: 

\begin{equation}
    \mathcal{L}_{\text{cyc}}(G_{A\shortrightarrow{P}}, G_{P\shortrightarrow{A}}) =  \mathbb{E}_{x_a\sim X_A} [\norm{x_a'-x_a}_1] 
    +  \mathbb{E}_{x_p\sim X_P}[\norm{x_p'-x_p}_1],
\end{equation}
where $x_a'=G_{P\shortrightarrow{A}}(G_{A\shortrightarrow P}(x_a))$ and $x_p'=G_{A\shortrightarrow{P}}(G_{P\shortrightarrow A}(x_p))$, $\norm{\cdot}_1$ is the $l_1$ norm.

The generative adversarial training with cycle-consistency enables synthesizing realistic looking radiographs across domains. However, there is no guarantee that high-level semantics would be preserved during translation. For example, when translating an adult X-ray with lung opacities, sometimes it might be converted into a normal pediatric X-ray without opacities, since the disease semantics are not explicitly modelled in the learning process.

\subsection{High-level class-specific semantics modelling}
To preserve high-level class-specific semantic information indicating abnormalities in the image before and after translation, we propose to explicitly model disease labels into the translation framework by incorporating auxiliary classification models with source labels. 

A source classification model $\mathcal{F}_A$ is first learned on the labeled source data $A=\{X_A, Y_A\}$ using a cross-entropy loss to classify $C$ categories:

\begin{equation}
    \mathcal{L}_{\text{cls}}(\mathcal{F}_A, A) = -\mathbb{E}_{a \sim A}\sum_{c=1}^C \mathbbm{1}_{c} \log \left( \sigma(\mathcal{F}_A^{(c)}(x_a)) \right), 
\end{equation}
where $\sigma$ is the softmax function, $\mathbbm{1}_c=1$ if an input image $x_a$ belongs to class $c\in C$, otherwise $\mathbbm{1}_c=0$. We then enforce the learned $\mathcal{F}_A$ to perform similarly on the reconstructed source data $A'=\{G_{P\shortrightarrow{A}}(G_{A\shortrightarrow P}(X_A)), Y_A\}$ to minimize $\mathcal{L}_{\text{cls}}(\mathcal{F}_A, A')$. In this way, the high-level class specific content is preserved within the \textit{source $\shortrightarrow$ target $\shortrightarrow$ source} cycle.

To retain similar semantics within the \textit{target $\shortrightarrow$ source $\shortrightarrow$ target} cycle in the absence of target labels $Y_P$, we learn a target classification model $\mathcal{F}_P$ (fine-tuned from $\mathcal{F}_A$) on synthetic target images to minimize $\mathcal{L}_{\text{cls}}(\mathcal{F}_P, \{G_{A\shortrightarrow P}(X_A) ,Y_A\})$, in the mean time,
minimizing $\mathcal{L}_{\text{cls}}(\mathcal{F}_A, \{G_{P\shortrightarrow A}(X_P), \text{arg} \max(\mathcal{F}_P(X_P))\})$, so that classifiers in both domains produce consistent predictions to keep semantic consistency. The total \textit{semantic classification loss} is:

\begin{align}
    \mathcal{L}_{\text{cls}}(\mathcal{F}_A, \mathcal{F}_P) & = \mathcal{L}_{\text{cls}}(\mathcal{F}_A, A) + \mathcal{L}_{\text{cls}}(\mathcal{F}_A, A') + \mathcal{L}_{\text{cls}}(\mathcal{F}_P, \{G_{A\shortrightarrow P}(X_A) ,Y_A\}) \nonumber \\
    & + \mathcal{L}_{\text{cls}}(\mathcal{F}_A, \{G_{P\shortrightarrow A}(X_P), \text{arg} \max(\mathcal{F}_P(X_P))\}). 
\end{align}

By modelling disease labels into the translation network, the synthesized images maintain meaningful semantics to favor the target clinically relevant task. For instance, $\mathcal{F}_P$ can be acted as a disease classifier on the target domain.

\subsection{Mid-level feature regularization}

Now that we have both low-level content and high-level semantics preserved in the transformation network, we further add mid-level feature constraints on the target model $\mathcal{F}_P$. This is done so that features extracted from the middle layers of $\mathcal{F}_P$ on real target data will be similar with that on synthetic target data.
Inspired by the perceptual loss~\cite{perceptual} that encourages image before and after translation to be perceptually similar, we impose \textit{feature reconstruction loss}, to encourage real target image $X_P$ and synthetic target image $G_{(A\shortrightarrow P)}(X_A)$ to be similar in the feature space. Using this feature regularization in training for middle layers of CNNs also tends to generate images that are visually indistinguishable from target domain referring to our experiments. The feature reconstruction loss is the normalized Euclidean distance between feature representations:

\begin{equation}
    \mathcal{L}_{\text{feat}}(\mathcal{F}_P) = \sum_i
  \frac{\|f_i - \hat f_i\|_2^2}{H_iW_iC_i}, 
\end{equation}
where $i$ is a convolutional block from target model $\mathcal{F}_P$, and $f_i$ and $\hat f_i$ are features maps of size $H_i \times W_i \times C_i$ output by the $i^{th}$ convolutional block.

\subsection{Final objective and implementation details}

The final objective of TUNA-Net is the sum of adversarial learning losses, cycle-consistency loss, semantic classification loss and feature reconstruction loss:

\begin{align}
    \mathcal{L} & = \mathcal{L}_{\text{adv}}(G_{A\shortrightarrow{P}},D_P) + \mathcal{L}_{\text{adv}}(G_{P\shortrightarrow{A}},D_A) \nonumber \\ & + \lambda\mathcal{L}_{\text{cyc}}(G_{A\shortrightarrow{P}}, G_{P\shortrightarrow{A}}) 
     + \mathcal{L}_{\text{cls}}(\mathcal{F}_A, \mathcal{F}_P) + \mathcal{L}_{\text{feat}}(\mathcal{F}_P). 
    \label{eq_final}
\end{align}

Driven by the target task of disease recognition, this corresponds to optimizing the objective for the adapted target model $\mathcal{F}_P$.

We adopt Cycle-GAN~\cite{CycleGAN2017} for training the I2I translation framework. We use 9 residual blocks~\cite{He_CVPR2016} for the generator network for an input X-ray image of size 512 $\times$ 512. For source classification networks $\mathcal{F}_A$, we use ImageNet pre-trained ResNet with 18 layers~\cite{He_CVPR2016} as a trade-off between performance and GPU memory usage. The target classification model $\mathcal{F}_P$ is fine-tuned from the source model $\mathcal{F}_A$ hence has the same network structure with $\mathcal{F}_A$. Feature maps of \textit{conv}\_3 (56 $\times$ 56 $\times$ 128) and \textit{conv}\_4 (28 $\times$ 28 $\times$ 256) are extracted from $\mathcal{F}_P$ as mid-level feature representations to calculate the reconstruction loss. $\lambda$ in Eq.~\ref{eq_final} is set to 10 as in~\cite{CycleGAN2017}. All other networks are trained from scratch with a batch size of 1, an initial learning rate of 0.0002 for first 100 epochs and linearly decay to 0 in the next 100 epochs. All the network components are optimized using the Adam solver. The TUNA-Net is implemented using the PyTorch framework. All the experiments are run on a 32GB NVIDIA Tesla V-100 GPU. 

\section{Experiments}
\label{exp}
\noindent\textbf{Material and settings:}
We extensively evaluate the proposed TUNA-Net for unsupervised domain adaptation on two public chest X-ray datasets containing normal and pneumonia frontal view X-rays, i.e., an adult chest X-ray dataset used in the RSNA Pneumonia Detection Challenge~\footnote{\url{https://www.kaggle.com/c/rsna-pneumonia-detection-challenge/data}} (a subset of the NIH Chest X-ray 14~\cite{Wang_CVPR2017}) and a pediatric chest X-ray dataset~\footnote{\url{https://doi.org/10.17632/rscbjbr9sj.3}} from Guangzhou Women and Children's Medical Center in China. We set the adult dataset as \textbf{source} domain and the pediatric dataset as \textbf{target} domain. For the adult dataset, we use 6993 normal X-rays and 4659 X-rays with pneumonia. For the pediatric dataset, we use 5232 X-rays (either normal (n=1349) or abnormal with pneumonia (n=3883), but labels were removed in our setting) for training and validation. The combined dataset are used to train the adult $\leftrightarrows$ pediatric translation framework. 5-fold cross-validation is performed.  Classification performance of the proposed adaptation method is evaluated on a hold-out test test of 624 pediatric X-rays (normal: 234, pneumonia: 390) from the target domain.

\noindent\textbf{Reference methods:}
Although unsupervised adversarial domain adaptation methods exist in medical imaging field, they are mainly designed for segmentation. Here we compare the performance of our proposed TUNA-Net with the following five relevant reference models:

1. \textbf{NoAdapt}: A ResNet-50~\cite{He_CVPR2016} CNN trained on adult X-rays is applied to the pediatric X-rays for pneumonia prediction. This serves as a lower bound method.

2. \textbf{Cycle-GAN}~\cite{CycleGAN2017}: Without considering labels indicating diseases in X-rays during I2I translation using~\cite{CycleGAN2017}. A model trained on labeled real adult X-rays is applied to synthetic adult X-rays generated from pediatric X-rays.

3. \textbf{ADDA}~\cite{ADDA_CVPR17}: First we train an adult classification network with labeled X-rays. Then we adversarially learn a target encoder CNN such that a domain discriminator is unable to differentiate between the source and target domain. During testing, pediatric images are mapped with the target encoder to the shared feature space of the source adult domain and classified by the adult disease classifier.

4. \textbf{CyCADA}~\cite{Cycada2017}: It improves upon ADDA by incorporating cycle consistency at both pixel and feature levels. 

5. \textbf{Supervised}: We assume that disease labels for target domain are accessible. A supervised model can be trained and tested on labeled target domain. This servers as an upper bound method.
\vspace{2mm}

\noindent\textbf{Quantitative results and ablation studies:}
We calculate the Area Under the Receiver Operating Characteristic Curve (AUC), accuracy (Acc.), sensitivity (Sen.), specificity (Spec.) and F1 score to evaluate the classification performance of our model. The validation set is only used to optimize the threshold using Youden's index (i.e., $max(\text{Sen.}+\text{Spec.}-1)$) for normal versus pneumonia classification. The classification results of our TUNA-Net and reference methods are shown in Table~\ref{results_comp}. The baseline method without adaptation (NoAdapt) performs poorly on the target task of pediatric pneumonia recognition, though the source classifier excels in pneumonia recognition on adult chest X-rays (AUC=98.0\%). It demonstrates that the gap between the source and target domain are fairly large although they share the same disease labels. Cycle-GAN does not consider disease labels during I2I translation. It generates X-rays without preserving high-level semantics, resulting in many normal adult X-rays converted into pediatric X-rays with opacities on the lungs, or adults with lung opacities converted into normal pediatric X-rays. This hugely decreases the adaptation performance for the classification task, where correct labels are considered to be crucial. Our full TUNA-Net considers high-level class-specific semantics achieves an AUC of 96.3\% with both sensitivity and specificity larger than 91\%. It outperforms both ADDA and CyCADA with similar settings. It is also worth noting that the performance of TUNA-Net is very close to that of the supervised model, where labeled training images on the target dataset are available. We ablate different modules in the TUNA-Net to see their influence on the final model: a). We exclude the feature construction loss in the target classification model; b). We do not use reconstructed images to retrain the source classification model; c). We exclude the target classification model $\mathcal{F}_P$ in the training, but use the synthetic images to train it offline. As shown in Table~\ref{results_comp}, each component contributes to improving the final TUNA-Net. The online end-to-end learning of $\mathcal{F}_P$ with other components is crucial and contributes most to the performance improvement. 
%%%%%%%%%%%%%%%%%%%%%%%%%%%%%%%Table%%%%%%%%%%%%%%%%%%%%%%%%%%%%%%%%%%%%%%%%%%%%%%%%%%%
\begin{table}[t]
  \centering
  % \scriptsize
  \normalsize
  \setlength{\tabcolsep}{5pt}
  \renewcommand{\arraystretch}{1.1}
  \caption{
    Comparison of normal versus pneumonia classification results on the test set of pediatric X-ray dataset.}
  \begin{tabular}{lccccc}
    \hline
	Model & AUC(\%) & Acc.(\%) & Sen.(\%) & Spec.(\%) &F1 \\
	\hline
	\hline
	NoAdapt
	&89.3$\pm$0.4 &82.5$\pm$0.3 &83.6$\pm$0.7 &80.8$\pm$0.8 &0.86$\pm$0.02\\
	Cycle-GAN~\cite{CycleGAN2017}
	&80.4$\pm$2.5 &74.2$\pm$2.7 &76.9$\pm$3.3 &69.9$\pm$2.8 &0.76$\pm$0.04\\
	\hline
	ADDA~\cite{ADDA_CVPR17}
	&91.8$\pm$0.4 &	88.1$\pm$0.4 &88.2$\pm$0.5 &87.0$\pm$0.4 &0.89$\pm$0.02\\
	CyCADA~\cite{Cycada2017}
	&93.5$\pm$0.5 &	90.0$\pm$0.4 &90.4$\pm$0.4 &89.6$\pm$0.5 &0.91$\pm$0.02\\
	\hline
	\textbf{TUNA-Net}
	&\textbf{96.3}$\pm$0.2 &\textbf{93.1}$\pm$0.4 &\textbf{92.9}$\pm$0.3 &\textbf{91.1}$\pm$0.4 &\textbf{0.93}$\pm$0.01\\

     a) w/o feature loss
     &95.9$\pm$0.1 &91.9$\pm$0.3 &91.7$\pm$0.3 &90.6$\pm$0.2 &0.92$\pm$0.01\\
     
     b) w/o $\mathcal{F}_A$ on rec.
     &94.6$\pm$0.2 &91.3$\pm$0.2 &91.0$\pm$0.3 &91.1$\pm$0.3 &0.92$\pm$0.01\\
     
     c) w/o $\mathcal{F}_P$, offline
     &94.1$\pm$0.2 &90.7$\pm$0.2 &91.0$\pm$0.4 &90.5$\pm$0.2 &0.91$\pm$0.01\\
     
	\hline
	\hline
	Supervised 
	&98.1$\pm$0.1 &96.3$\pm$0.1 &94.6$\pm0.3$ &92.8$\pm0.2$ &0.96$\pm0.01$\\
    \hline
  \end{tabular}
  \label{results_comp}

\end{table}
%%%%%%%%%%%%%%%%%%%%%%%%%%%%%%%Table%%%%%%%%%%%%%%%%%%%%%%%%%%%%%%%%%%%%%%%%%%%%%%%%%%%

%%%%%%%%%%%%%%%%%%%%%%%%%%%%%%%Fig%%%%%%%%%%%%%%%%%%%%%%%%%%%%%%%%%%%%%%%%%%%%%%%%%%%
\begin{figure*}[t!]
  \centering
  \includegraphics[width=\linewidth]{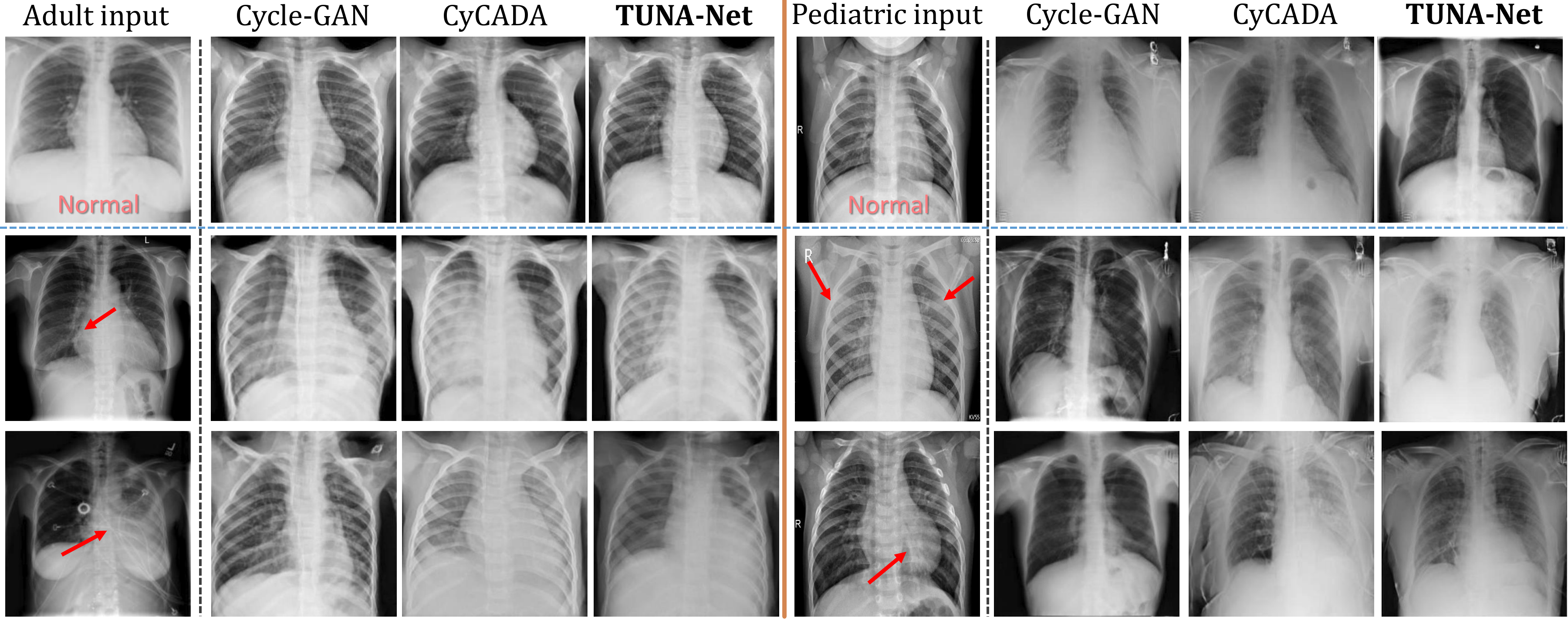}
  \caption{Qualitative comparison of image-to-image translation. Cycle-GAN is trained without using labels indicating normal or pneumonia, while CyCADA and our TUNA-Net considers labels in source domain in training. Left part shows adult $\shortrightarrow$ pediatric, right shows pediatric $\shortrightarrow$ adult. The first row shows two normal X-rays as input. The appearances of pneumonia(s) are pointed by arrows. Please refer to supplementary material for higher resolution images.}

  \label{fig:results}
\end{figure*}
%%%%%%%%%%%%%%%%%%%%%%%%%%%%%%%%%%%%%%%%%%%%%%%%%%%%%%%%%%%%%%%%%%%%%%%%%%%%%%%%%%%%%%

\noindent\textbf{Qualitative results:}
We show some qualitative image-to image translation examples in Figure~\ref{fig:results}. Cycle-GAN failed to preserve important semantic information during transfer. CyCADA is able to preserve certain high-level semantics but not as robust as the proposed TUNA-Net. TUNA-Net retains image content of various levels: from low-level content, mid-level features, to high-level semantics. For example, for the bottom left adult input, Cycle-GAN removes the pathology while our TUNA-Net perfectly preserves it. The synthetic X-rays by TUNA-Net are most close to the input source image semantically and to the target domain anatomically. 

\noindent\textbf{Discussion:} We specifically focused on normal versus pneumonia classification on a cross-domain setting. We showed that the I2I translation framework can be constrained using semantic classification components to preserve class-specific disease content for medical image synthesis. We used two public chest X-ray datasets with sufficient domain shift to demonstrate the ability of our unsupervised domain adaptation method. The domain adaptation from adult to pediatric chest X-rays is natural and intuitive. For example, medical students and radiology residents learn in a similar way: they first learn to read adult chest X-rays, and then they transfer the learned knowledge to pediatric X-rays.

\section{Conclusion}

In this paper, we investigated how knowledge about class-specific labels can be transferred from a source domain to an unlabeled target domain for unsupervised domain adaptation. Using adversarially learned cross-domain image-to-image translation networks, we found clear evidence that semantic labels could be translated across medical image domains. The proposed TUNA-Net is general and has the potential to be extended to more disease classes (\eg, pneumothorax), other image modalites (such as CT and MRI) and more clinically relevant tasks.

\subsubsection*{Acknowledgments.} This research was supported by the Intramural Research Program of the National Institutes of Health Clinical Center and by the Ping An Technology Co., Ltd. through a Cooperative Research and Development Agreement. The authors thank NVIDIA for GPU donations.

\bibliographystyle{splncs04}
\bibliography{ref}

\end{document}